\documentclass[epsf]{aa}
\usepackage{graphicx}

\begin{document}
\def\fu{$f_1$}
\def\t{$\pm$}
\def\fd{$f_2$}
\def\ft{$f_3$}
\def\fq{$f_4$}
\def\fdu{$f_2 - 2f_1$}
\def\fp{$f_1 + f_2$}
\def\fm{$f_2 - f_1$}
\def\cd{cd$^{-1}$}
\def\cds{cd$^{-1}$\,}
\def\kms{km~s$^{-1}$}
\def\kmss{km~s$^{-1}$\,}
\def\I{\'\i}
\def\salp{\vskip 0.3truecm}
\title{The multiperiodicity of the $\gamma$ Doradus stars HD~224945 and HD~224638 as
detected from a multisite campaign\thanks{Based on
observations partially collected at ESO-La Silla (Proposals 54.E-018 and 56.E-0308)}}
\author{E.~Poretti\inst{1}, C.~Koen\inst{2}, M.~Bossi\inst{1},
E.~Rodr\'{\i}guez\inst{3}, S.~Mart\'{\i}n\inst{3,1},
K.~Krisciunas\inst{4}, M.C.~Akan\inst{5}, R.~Crowe\inst{6},
M.~Wilcox\inst{6}, C.~Ibanoglu\inst{5}, S.~Evren\inst{5}}
\institute {Osservatorio Astronomico di Brera, Via Bianchi 46,
I-23807 Merate, Italy\\
 \email{poretti@merate.mi.astro.it}
\and South African Astronomical Observatory, PO Box 9, Observatory 7935, Cape Town, 
South Africa
\and Instituto de Astrofisica de Andalucia, C.S.I.C., Apdo. 3004, 18080 Granada, Spain
\and Cerro Tololo Inter-American Observatory, Casilla 603, La Serena, Chile
\and Ege University, Science Faculty, Dept of Astronomy and Space Sciences,
   Bornova 35100 Izmir, Turkey
\and Department of Physics and Astronomy, University of Hawaii $-$ Hilo, 200 West Kawili Street,
   Hilo, Hawaii 96720-4091, USA
}
\offprints{E. Poretti}
\date{Received date; Accepted Date}
\abstract{We discuss new photometric data collected on the $\gamma$ Dor
variables HD~224945 and HD~224638. Multiperiodicity was detected 
in both stars, thanks to the clear spectral window of a multisite campaign
that involved five observatories. HD~224945 shows the shortest period among
the $\gamma$ Dor stars, i.e., 0.3330~d. The
pulsation behaviour is very different: HD~224945 displays a set of
frequencies spread over an interval much wider than that of HD~224638.
We clearly found evidence for amplitude variations in the excited modes by
comparing data from different years. HD~224945 and HD~224638 are among the
best examples of $\gamma$ Dor stars that show multimode pulsations, which make
them very interesting from an asteroseismological point of view.
}
\authorrunning{Poretti et al.} 
\titlerunning{HD 224945 and HD 224638}
\maketitle

\section{Introduction} 
The variability of HD 224638$\equiv$BT Psc ($V$=7.5, F1~V) and 
HD 224945$\equiv$BU Psc ($V$=6.93, F0~V)
was announced by Mantegazza \& Poretti
(1991), as a by--product  of the monitoring of the $\delta$ Sct star HD
224639$\equiv$BH
Psc. Both stars had been used as comparison stars in the 
first observing run devoted to BH Psc. They increased the number of known 
$F$--type
stars located close to the low--temperature edge of the Cepheid instability strip
which exhibit small amplitude variability on time scales of several hours, usually
longer than the length of a night of observation
and therefore easily detectable only when used as comparison stars for short--period
variables. 
At that time, the debate on the nature of these variations was divided between 
spot activity (observed periods as rotational periods) and pulsation
(nonradial $g$--modes). 
Mantegazza et al. (1994, hereinafter Paper~I)
tried to explain the complicated light behaviour of HD 224639 and HD 224945 in the
simplest way possible, by means of periodicities shaped as double-- or triple--wave
curves.  Balona et al. (1994) reported on
the multiperiodicity of $\gamma$ Dor, giving decisive evidence in favour of
pulsation. Also, as a result of combined photometry and line profile variations
for 9 Aurigae (Krisciunas et al. 1995) and $\gamma$ Dor (Balona et al. 1996), plus
the preliminary results on HD 224945 (Poretti et
al. 1996), the  hypothesis of variability caused by $g$--modes  achieved 
a wide consensus. The properties of this new  class of variable stars were
delineated step--by--step through observational efforts and have been summarized by
Kaye et al. (1999),
while the problem of the driving mechanism and the excitation of $g$--modes
constitutes the target of continuing theoretical investigations.
As a result,
$\gamma$ Dor variables are now  considered intermediate between $A-F$ pulsators and
solar--type stars, finding  a special place  in the main programmes of 
asteroseismologic missions such as {\sc corot} and {\sc mons}. 

 On the basis of this progress,
it is worth investigating in more detail the pulsational behaviour of HD 224639 
and HD 224945. Here we present the results of 
a multisite campaign carried out in October 1995.

\section{Observations}

HD 224945 and HD 224638 are located very close to the celestial equator and therefore
can be monitored from both the northern and southern hemispheres. As comparison stars we
used the same
two stars as in the 1991 campaign (HD 225086 and HD 200), since they
proved to be stable within a few mmag. Five telescopes were used for this
multisite campaign: 
\begin{enumerate}
\item The European Southern Observatory  50--cm telescope located in 
La Silla (Chile), equipped
with a photon--counting photometer (EMI~9789 QB photomultiplier) and
$B$ and $V$ filters. The observer was
E.~Poretti. 
\item The 48--cm telescope, located at Ege University Observatory, was 
equipped with a solid--state photometer and $B$ and $V$ filters. The observer 
team was led by M.~C.~Akan.
\item The 50--cm telescope located in Sutherland, South African Astronomical Observatory,
was equipped with a photon--counting photometer and $B$ and $V$ filters.
The observer was C. Koen.
\item The 61--cm telescope located on the  Mauna Kea (Hawaii) was equipped
with  a photon--counting photometer and standard $B$ and $V$ filters. The
observers were K.~Krisciunas, R.~Crowe and M.~Wilcox.
\item The 90--cm telescope located at Sierra Nevada Observatory (Spain).
The observers were E. Rodr\'{\i}guez and S. Mart\'{\i}n. Photometry was performed in
the $uvby$ system, but only $v$ and $y$ measurements are discussed here, as
the more compatible with the $B$ and $V$ ones, respectively.
\end{enumerate}

Earlier observations were performed in 1994 at European Southern Observatory,
using the ESO  50--cm telescope equipped with the same instrumentation
used in 1995. On that occasion the observer was M.~Bossi; the results obtained
on BH Psc are reported by Mantegazza et al. (1996). The measurements were
distributed over 13 nights, with a total useful observing
time of 64 hours and a baseline of 14.9 d.

\begin{figure*}
\resizebox{\hsize}{!}{\includegraphics{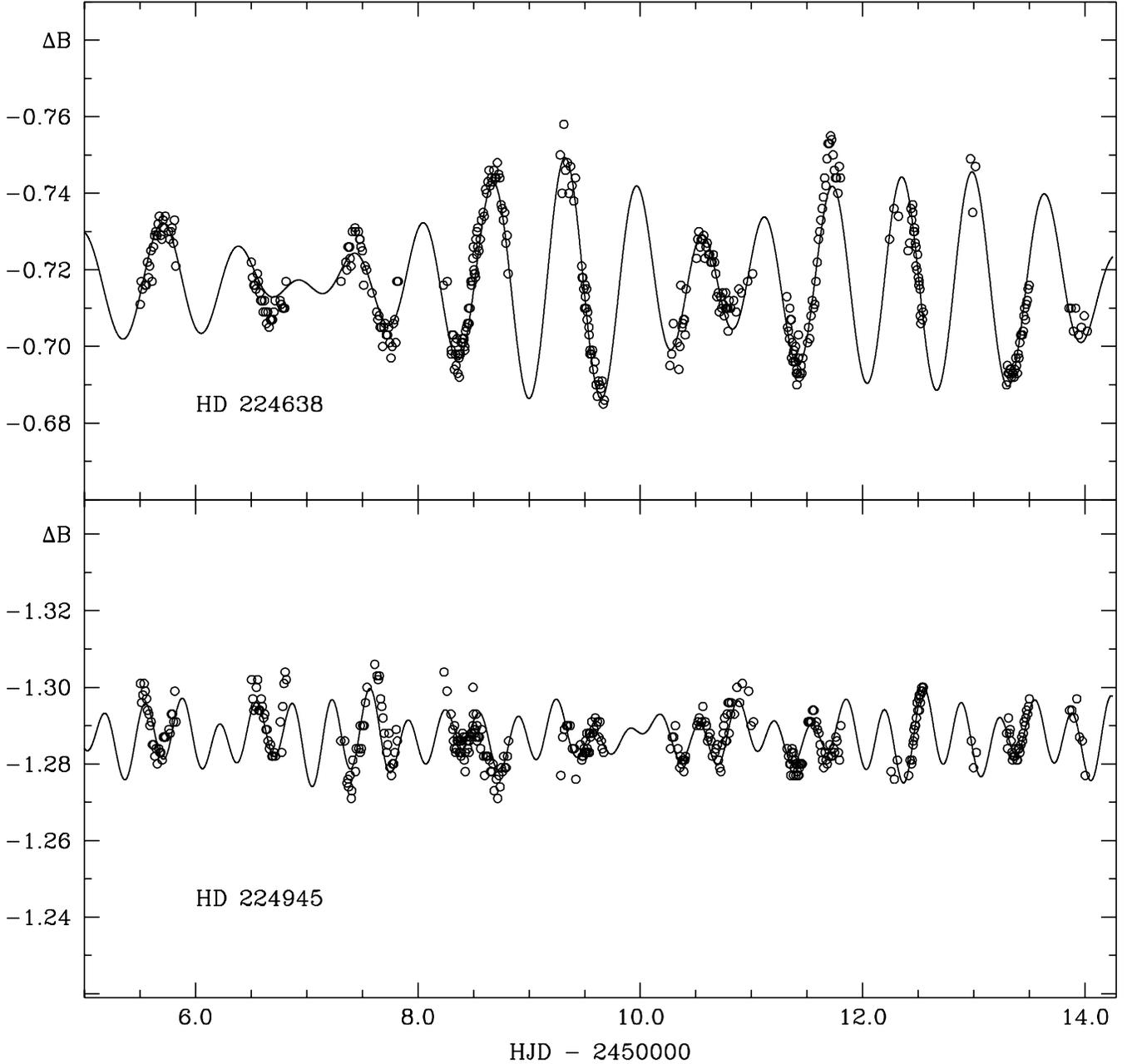}}
\caption[ ]{Light curves ($B$ light) of HD~224945 and HD 224638 obtained during the
1995 multisite campaign; measurements from JD 2450005 to 2450014 are shown. 
Note the different timescales and amplitudes between
the two curves.}
\label{lc}
\end{figure*}

\section{Data reduction and analysis}
The original measurements were provisionally reduced by the observers themselves
to check the data quality and for comparison purposes. The original
raw data were then re--reduced by
means of the same algorithm; this homogenous procedure allowed us to avoid
the introduction of local drifts caused by  routines using different
interpolation formulae and/or different methods of applying extinction
corrections. Since the weather was not particularly favourable during
the campaign, the rejection of some parts of nights characterized by
unfavourable weather conditions was decided on the basis of a uniform
criterion. We also decided to use the code which allows us to calculate 
instantaneous values of the extinction coefficients (Poretti \& Zerbi 1993)
since in some cases changes were expected to occur.

Table 1 lists some details regarding
observing runs at each site. As can be noted, the mean values of the
magnitude differences between the two comparison stars show the presence of
systematic shifts (up to 0.03 mag). This is not surprising, taking
into account that different photomultipliers and filters  were used.
The systematic shifts between HD~225086 and HD~200 cannot be
used to correct magnitude differences between HD~225086 and the two
program stars since they have different colours from HD~200; uncertainties
of the colour transformations are larger than the precision we need. 
Hence, to apply this correction 
we exploited the very useful circumstance that the spread in longitude between
different observing sites  is not
larger than the length of the night.  There were always some overlapping segments 
of the light curve, which allowed us to 
determine the amounts of the corrections in a straightforward way.

It should also be noted that the Turkish measurements show a scatter larger
than all the others.  At first, all the Turkish data were considered, but
in the last step of the analysis (once it was established that rapid variations are
not present in the light curves) those measurements were averaged in groups of 
4--5. Thus, the final set of data contains the original measurements from
all the observatories except for the averaged values from Turkey.
The standard deviations of the $\Delta B$ and $\Delta V$ values 
between the comparison stars are 3.7 and 3.8 mmag, respectively.
\begin{table*}
\begin{flushleft}
\caption{Summary of the photometric $B$ and $V$ data collected in the
1995 multisite campaign. $N$ is the total
number of measurements, $n$ the number of nights. $t_1$
and $t_N$ are the times of the first and last measurement,
respectively. Magnitude differences are calculated with respect to
HD~225086.}   
\begin{tabular}{l c rrr c rrrr c rrrr}
\hline
\hline
\multicolumn{1}{c}{Site}& &\multicolumn{3}{c}{Check star (HD 200)} & 
&\multicolumn{4}{c}{HD 224945}& & \multicolumn{4}{c}{HD 224638}\\
 & & $\Delta$V & N & s.d. & & t$_1$ & t$_{\rm N}$ & n & N  & &
t$_1$ & t$_{\rm N}$ & n & N \\
\noalign{\smallskip}
\hline
\noalign{\smallskip}
ESO       & & 0.1472 & 198 & 0.0029 & & 4.605 & 14.597 & 10 & 205  & 
                                      & 4.512 & 14.572 & 10 & 219  \\
Ege Obs.  & & 0.1180 & 117 & 0.0102 & & 3.250 & 12.317 & 5  &  36  &
                                      & 3.252 & 12.311 & 5 &   33    \\
SAAO      & & 0.1294 & 41  & 0.0055 & & 1.427 & 14.479 & 6  &  36  &
                                      & 1.432 & 14.482 & 6 &   44  \\
Mauna Kea & & 0.1371 & 39  & 0.0051 & & 7.824 & 14.003 & 6  &  37  &
                                      & 7.840 & 14.023 & 6  &  38  \\
OSN       & & 0.1304 & 126 & 0.0032 & & 7.364 & 19.524 & 9  & 128  &
                                      & 7.362 & 19.527 & 9  & 133  \\
\noalign{\smallskip}
\hline
\noalign{\smallskip}
 & & $\Delta$B & N & s.d. & & t$_1$ & t$_{\rm N}$ & n & N  & &
t$_1$ & t$_{\rm N}$ & n & N \\
\noalign{\smallskip}
\hline
\noalign{\smallskip}
ESO       & & 0.2569 & 198 & 0.0029 & & 4.604 & 14.597 & 10 & 203  & 
                                      & 4.512 & 14.572 & 10 & 217  \\
Ege Obs.  & & 0.2672 & 125 & 0.0091 & & 3.264 & 12.312 & 5  &  32  &
                                      & 3.251 & 12.322 & 5  &  29\\
SAAO      & & 0.2458 & 41  & 0.0060 & & 1.436 & 14.479 & 6  &  32  &
                                      & 1.432 & 14.482 & 6  &  44 \\
Mauna Kea & & 0.2554 & 20  & 0.0041 & &10.751 & 14.002 & 3  &  21  &
                                      &10.743 & 14.021 & 3  &  21 \\ 
OSN       & & 0.2449 & 126 & 0.0029 & & 7.364 & 19.524 & 9  & 129  &
                                      &  7.362 & 19.527 & 9  & 133  \\
\hline
\hline
\end{tabular}
\end{flushleft}
\end{table*} 

Figure~\ref{lc} shows a part of the $B$ light curves of HD 224945 and HD 224638. The fits
derived in the next sections are also shown. The original measurements can be
requested from the authors.

Since data in the final form circulated amongst all the
participants, several co--authors  analysed the time series
independently and by different period search algorithms 
(least--squares, {\sc clean, dft},~...)
and it was particularly satisfying to see that they 
detected the same terms, even if, owing to the complexity of the light 
variations, the terms did not always stand out in a clear way. 

We present here the analysis carried out by
using the least--squares method (Vani\^cek 1971) used by the Merate group
in the analysis of $\delta$ Sct star light curves. This method has the
advantage of not using any data prewhitening since only the frequency values
previously found are considered as input values (known constituents; k.c.'s);
their amplitude and phase values are recalculated as unknowns in the new searches.
That means that after the detection of the \fu~term, only the frequency value
\fu was considered as established (i.e. a k.c.) and in the second search the
unknowns were $\Delta m_o, A_1, \phi_1, f_2, A_2, \phi_2$. The ordinates of the
power spectra show the reduction factor
\begin{equation}
{\rm Red. Factor} = 1 - \frac{\sigma^2_{\rm fin}}{\sigma^2_{\rm in}}
\end{equation}

Moreover, we can present the term detection step--by--step. The
frequency values were refined after each new detection.
We can also obtain useful evidence concerning the amplitude by applying a
least--squares
fit to the datasets. The interpolating formula we used is 
\begin{equation}
\Delta m(t)= \Delta m_o + \sum_i^N {A_i \cos [2\pi f_i  (t-T_o) +\phi_i ]}
\end{equation}

Since the 1995 observations are concentrated in an interval of 15~d only,
the frequency resolution is 1.5/$\Delta$T=0.10~\cd. However, the absence of relevant
aliases makes the detection of the excited terms very simple.
We note that the frequencies mentioned 
in the text and figures are given to better accuracies that the formal 
frequency resolution, because they were determined by a least--squares
procedure. The formal least--squares standard errors on the 
frequencies no doubt understimate the true uncertainties -- aside from
problems pointed out by e.g. Montgomery \& O'Donoghue (1999), it is clear
that we have not always resolved close frequencies, and that some unidentified
signals may remain in the data.

\section{Frequency analysis of the HD 224945 data}
\subsection{The 1995 campaign} 
\begin{figure}
\resizebox{\hsize}{!}{\includegraphics{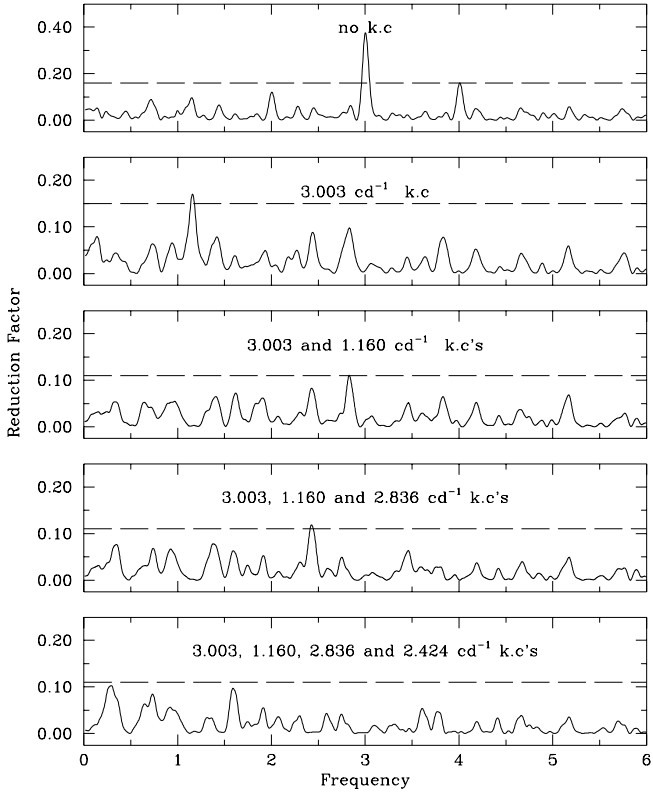}}
\caption[ ]{Power spectra of the $B$ measurements of HD~224945 obtained during the
1995 multisite campaign. Note the different scale of the top panel.}
\label{sp95}
\end{figure}

We collected 442 $V$ and 417 $B$ measurements of HD~224945, considering
the Turkish measurements as binned points. The results
of the frequency analysis of the $B$ data are shown in Fig.~\ref{sp95}.
In each panel the horizontal line indicates the level for 
$S/N=4.0$, i.e., the limit usually accepted for significance
(Kuschnig et al. 1997).

The top panel shows the spectrum obtained without any k.c.: the peak
at 3.00~\cds~ stands out clearly, and the alias structure mimics very well
the spectral window (see also Fig.~1 in Poretti et al. 1996). In the 
discussion of the 1991 data (Paper~I),  
the alias at 2.00~\cds was erroneously preferred. 
The reality of this term close to a multiple of 1~\cds was
discussed in Paper~I.  
We note that it is also detected in the new campaign data for
HD~224945, but in neither the HD~224638, nor the HD~200, time series;
its physical presence in the HD~224945 data is therefore certain. 
When introducing the 3.00~\cds frequency  
as k.c., the power spectrum shows an almost flat pattern,
indicating that the amplitude of the remaining terms is smaller.
However, two peaks reach the acceptance level: the first is located at   
1.16~\cds (second panel), the second at 2.84~\cds (third panel).
After introducing these three terms as k.c.'s, the detection of
further terms becomes delicate. The highest peak in the fourth
panel is at 2.42~\cds and after that the power spectrum does not
show a really dominant peak (bottom panel). 

 The analysis of the
$V$ measurements yields the same results, except with a slight
difference in the frequency of the 2.84~\cds term. A value close
to 2.77~\cds is preferred.  Since the peak at 2.84~\cds is broad 
(see third panel of Fig.~\ref{sp95}), the presence of another
undetectable term 
combined with the  noise distribution could  be responsible for the
difference between the two values. In the least--squares solution we will consider
the 2.84~\cds term as the only well--established term.  

These independent terms describe a different scenario from that  
in Paper~I. In that case the single--site
observations make the power spectrum very complicated, with a great
uncertainty between a peak and its $\pm$1~\cds aliases. This
uncertainty led us to propose a solution based on two periodicities
each having a triple--wave shape, as it was considered the 
simplest. Thanks to the multisite observations,
we now know that  many pulsational modes are simultaneously excited. 

\subsection {The 1994 and 1991 observations}
The frequency analysis of the $B$ data collected in the 1994
season  is much more complicated 
than that of the multisite campaign. In spite of that, we could 
detect the 3.00 and the 2.84~\cds terms.
When introducing these two terms as k.c.'s,
the highest peak is at 2.27~\cd. This peak is significantly
different than the 2.42~\cds detected in the 1995 multisite campaign.
The search for a fourth component is
not easy since we get numerous peaks: however, we could recognize the
1.16~\cds term and its aliases. We considered it as a real mode on the
basis of the results of the 1995 campaign. After that, 
the new search showed evidence of a complicated structure
centered at 1.66~\cd. We have no way
of selecting in a reliable
way the true term amongst this peak and the aliased ones. Moreover,
the S/N of this term is less than 3.0. 
The $V$ data allowed us to identify the same terms, supporting the
presence of the new term at 2.27~\cds. 
 
In the 1991 dataset the detection
order of the terms is 3.00, 2.84 and 2.31~\cds: 
the only difference with respect to the previous least--squares
search (see the beginning of \S~3.2 in Paper~I)
is the identification of the first term as 3.00~\cds instead of 2.00~\cds.
No other term can be detected after considering this triplet as k.c.'s;
in particular there is no evidence of the presence of the 1.16~\cds term. 

It should be emphasized that it would be very difficult to solve the 1994
and 1991 light curves without having  the way shown by the results
of the 1995 campaign. In turn, the similar structures observed in
the three independent datasets corroborate the proposed identification of
the excited frequencies, even if some of them are at the limit of the acceptance level.

\subsection{Frequency refinement and least--squares fitting}

The much longer time interval covered by
the 1991 data (the two runs span 14 and 10 days, respectively,
and they are separated by a gap
of 10 days) allowed us to use these data to refine the frequency values. 
First, we calculated
the best solution using all the 1991 data. Then 
we calculated the amplitudes for each 1991 subset keeping fixed frequencies and
phases. 

Before applying these frequency values to other datasets, we verified the 
correspondence between some terms.
We  considered the 2.27~\cds term detected
in the 1994 data as the same detected  at 2.31~\cds in the 1991 campaign.
However, we note that the least--squares solution yields a better fit when
considering different values for different seasons. On the other hand, we
cannot consider the 2.42~\cds term detected in 1995 as coincident with the
2.31~\cds term. We suggest that there are many terms excited (simultaneously
or not) in the narrow range 2.2--2.5~\cd and their detection is not an easy
task. 
We considered the following five terms as independent modes detected
in the three campaigns: 3.003, 2.836, 2.313, 2.424  and 1.160 \cd. 
We also note the almost 2:1 ratio between the 2.313 and 1.160~\cds terms: the similar
amplitude does not support a monoperiodic contribution to the light curve 
as it should be very
asymmetric. The resonance effect is more plausible, even if it should be noted
that the 1.160 \cds term was not observed in the 1991 dataset.

Using the four frequencies above, we calculated the amplitudes from the 1994 
and 1995 datasets.
Tab.~\ref{lsq45} summarizes the
results. The formal errors of the amplitudes are about 0.4 mmag.
Looking at the differences between the amplitudes of the two 1991 subsets
we can see that they are marginally significant, being in the interval 0.7--0.9
mmag. This very close similarity demonstrates that solutions
obtained from short runs are self--consistent. 
The dense time coverage in each dataset ensures the reliability of
the amplitude determination. Therefore, we can
look at the differences in the amplitudes between the 1991, 1994 and
1995 datasets with greater confidence. 

We also analyzed in frequency the data obtained by merging
the 1991, 1994 and 1995 datasets. For each detected term, it is
not possible to select the true value owing to the numerous aliases separated
by integer values of $\pm$1~cy$^{-1}$; therefore, in the case of
HD~224945 we cannot
propose a full, comprehensive solution and evaluate in a reliable way the
phase coherence of the pulsation. The amplitude variability is a
further complication.  
 It should be noted that an attempt at such an analysis
detected two peaks  at 2.27 and 2.31~\cd, suggesting  that these two peaks
are related to  two independent terms (see the above comparison of the 1994 and
1991 results). 

The  sum of the squared amplitudes  
are the same in 1991 and 1994 (61 and 58 mmag$^2$, respectively),
while it is lower in 1995 (50 mmag$^2$). However, the amplitude of  a mode
can change dramatically: the 2.836~\cds term halved its $B$ amplitude from
1991 to 1995 while the 2.313~\cds term disappeared.
As a matter of fact, in the 1991 dataset there are three terms having
the same amplitude, while in the 1995 ones the 3.003~\cds term is
dominating. The amplitude variability looks like a 
well--established fact for this $\gamma$ Dor star.

\begin{table*}
\begin{flushleft}
\caption{HD 224945: least--squares fit of the $B$ and $V$ measurements performed
in the 1991, 1994 and 1995 observing seasons. The term detected at 2.27~\cds in the 1994
season is considered the same as that detected at 2.31~\cds in the other seasons,
while that at 2.42~\cds is considered an independent frequency. Formal errors
on the frequencies are calculated with respect to the 1991 data (when applicable) or to
the 1995 data.}
\begin{tabular}{l c ccc c c c c c c c c}
\hline
\multicolumn{1}{c}{}& &\multicolumn{7}{c}{Amplitude $B$ [mmag]} & &
\multicolumn{3}{c}{Amplitude $V$ [mmag]}\\
\cline{3-9}\cline{11-13}\\
\multicolumn{1}{c}{Freq.}& &\multicolumn{1}{c}{1991}&
\multicolumn{1}{c}{1991}&\multicolumn{1}{c}{1991}&  
& \multicolumn{1}{c}{1994}& &\multicolumn{1}{c}{1995}& &
\multicolumn{1}{c}{1994}& &\multicolumn{1}{c}{1995}\\
\cline{3-5}
\multicolumn{1}{c}{[\cd]}& &\multicolumn{1}{c}{All}&
\multicolumn{1}{c}{I}&\multicolumn{1}{c}{II}&  
& \multicolumn{2}{c}{} &\multicolumn{2}{c}{}\\
\noalign{\smallskip}
\hline
\noalign{\smallskip}
3.003$\pm$0.001  & &  4.3 & 4.5 & 3.8  & &    5.4 & & 6.1 &  & 4.0  & & 5.1 \\
\noalign{\smallskip}
2.836$\pm$0.001  & &  4.9 & 5.2 & 4.3  & &    4.0 & & 2.5 &  & 3.4  & & 1.5 \\
\noalign{\smallskip}
2.313$\pm$0.001  & &  4.3 & 4.5 & 3.8  & &    2.5 & & -- &  & 2.3  & & --  \\
\noalign{\smallskip}
2.424$\pm$0.005   & &  -- & -- & --  & &    -- & &  2.3 &  & --  & & 1.9  \\
\noalign{\smallskip}
1.160$\pm$0.005   & &  --  & --  & --   & &    2.5 & & 3.0 &  & 2.2  & & 2.4 \\
\noalign{\smallskip}
Res. rms [mmag]  & &  5.4 & 5.3 & 5.4  & &    5.1 & & 4.2 &  & 5.0  & & 4.2 \\
\hline
\end{tabular}
\label{lsq45}
\end{flushleft}
\end{table*}

\section{Frequency analysis of the HD 224638 data}

\subsection{The 1995 campaign}
\begin{figure}
\resizebox{\hsize}{!}{\includegraphics{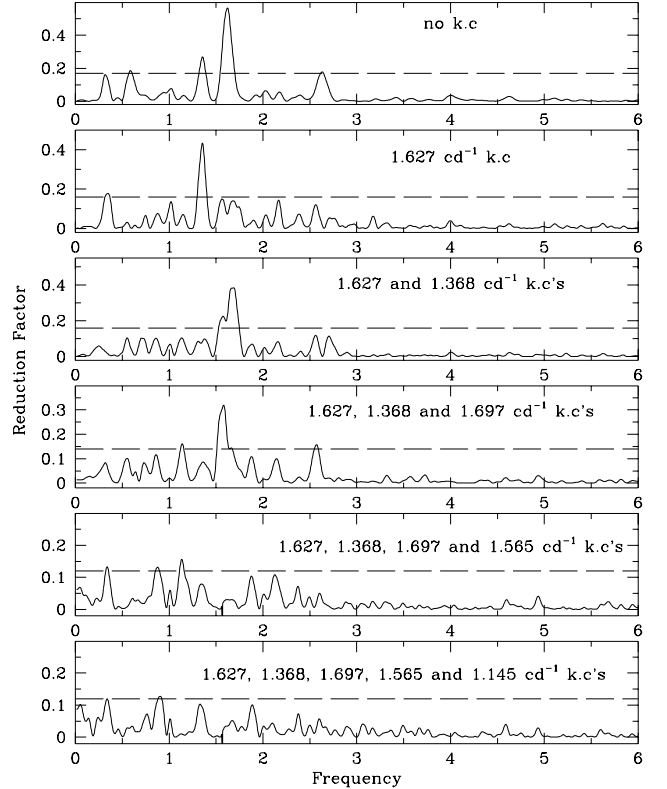}}
\caption[ ]{Power spectra of the $B$ measurements of HD 224638 obtained during the
1995 multisite campaign. Note the different scales of the panels.}
\label{sp3895}
\end{figure}

The analysis of the most recent $B$ and $V$ measurements of HD 224638 identified
the same terms as before,
i.e., 1.627, 1.368, 1.697, 1.565 and 1.145~\cds (Fig.~\ref{sp3895}). Also in this case
the multisite observations allowed us to identify the terms without any ambiguity.
Differently than the case of HD~224945, there are several very close frequencies:
1.627 (top panel), 1.697 (third panel), and 1.565~\cds (fourth panel). 
The separations between the two side peaks and the central one are 0.07 and 0.06~\cd;
these values are comparable to the half width at the half maximum, i.e. 
1/$\Delta T$=0.07~\cd. Indeed, in the power
spectra the peaks are not well resolved, but they appear to be double and/or 
enlarged.  In particular, the double peaks appearing in the third and fourth
panels of Fig.~\ref{sp3895}
should be noted.  This reveals the presence of overlapping peaks. Two peaks
flanking a
central one at the limit of the frequency resolution can be generated by
a single term  modulated in amplitude; however,  
the amplitudes of the side peaks are comparable to that of the central
one and this fact supports the reality of the triplet. The analysis
of the 1991 dataset definitely confirms the hypothesis of the three
independent modes (see next subsection). Note that the amplitudes
of these terms are larger than those detected in the HD~224945 data,
giving a higher $S/N$ value.  The other
two terms (1.368, second panel, and 1.145 \cd, fifth panel) are more separated.

The power spectrum obtained by introducing the five frequencies as k.c.'s is
not homogenously flat (bottom panel). The noise in the region 0.0--2.0~\cds 
is 1.34 mmag, while
in the region 4.0--6.0~\cds it is only 0.65 mmag. However, it should be noted that 
the general patterns of the residual $B$ and $V$ power spectra are
slightly different.
This fact suggests that noise dominates over pulsation
and the latter is really undetectable owing to the very small amplitude of the involved
terms.

\subsection{The 1994 and 1991 seasons}
Once again, the light curves of the 1991 and 1994 seasons can only be
understood if extra alias-free frequency information is available.
The frequency analysis of the 1991 $B$ data provided evidence for
the terms 1.697, 1.627,
1.368 and 1.565 \cd. The first three terms were already reported in Paper I, where
two periodicities having each the shape of a double--wave were proposed.
The 1.565~\cds (detected also in the 1995 dataset) is a new term obtained by 
pushing one further step in the analysis. As a result, we also detect here  
the close triplet observed in the 1995 campaign.  Since the frequency
resolution in the 1991 dataset is much better than in the 1995 one, this 
confirmation strengthens our confidence in the reality of the triplet.
The peak at 1.145~\cds does not appear
very clearly in the power spectrum, but it has been considered for
the least--squares solution.


The first four terms are also detected in the 1994 $B$ and $V$ datasets, but 
residual signal is left, especially in the
region 0.5--1.0~\cd. The 1.145~\cds term shows an amplitude much smaller than 
the other four terms. As a matter of fact, 
in this season it seems that many terms having amplitudes at
the mmag level are excited.

\begin{table*}
\begin{flushleft}
\caption{HD 224638: least--squares fit of the $B$ and $V$ measurements performed
in the 1991, 1994 and 1995 observing seasons. Formal errors on frequencies
are calculated  with respect to the 1991 data.}
\begin{tabular}{l c rrr c r c r c r c c}
\hline
\multicolumn{1}{c}{}& &\multicolumn{7}{c}{Amplitude $B$ [mmag]} & &
\multicolumn{3}{c}{Amplitude $V$ [mmag]}\\
\cline{3-9}\cline{11-13}\\

\multicolumn{1}{c}{Freq.}& &\multicolumn{1}{c}{1991}&
\multicolumn{1}{c}{1991}&\multicolumn{1}{c}{1991}&  
& \multicolumn{1}{c}{1994}& &\multicolumn{1}{c}{1995}& &
\multicolumn{1}{c}{1994}&  &\multicolumn{1}{c}{1995}\\
\cline{3-5}
\multicolumn{1}{c}{[\cd]}& &\multicolumn{1}{c}{All}&
\multicolumn{1}{c}{I}&\multicolumn{1}{c}{II}&  
& \multicolumn{2}{c}{} &\multicolumn{2}{c}{}\\
\noalign{\smallskip}
\hline
\noalign{\smallskip}
1.627$\pm$0.001   & &  9.1 & 9.0 & 8.9  & &    7.1 & & 10.8 &  & 6.2  & & 9.2 \\
\noalign{\smallskip}
1.368$\pm$0.001   & &  9.0 & 8.4 & 8.7  & &    7.3 & & 10.2 &  & 6.0  & & 7.4 \\
\noalign{\smallskip}
1.697$\pm$0.001   & &  8.0 & 7.1 & 9.4  & &    7.7 & &  7.4 &  & 6.6  & & 6.3  \\
\noalign{\smallskip}
1.565$\pm$0.001   & &  4.7  & 5.0  & 6.0   & &    9.1 & & 5.2 &  & 7.0  & & 4.0 \\
\noalign{\smallskip}
1.145$\pm$0.001   & &  2.1  & 2.6  & 4.5   & &    2.1 & & 3.0 &  & 1.7  & & 2.1 \\
\noalign{\smallskip}
Res. rms [mmag]   & &  6.4  & 4.9  & 7.1   & &    5.6 & & 4.7 &  & 5.1  & & 4.3 \\
\noalign{\smallskip}
\hline
\end{tabular}
\label{lsq38}
\end{flushleft}
\end{table*}

\subsection{Frequency refinement and least--squares fitting}
We followed the same procedure used for HD 224945, refining the frequency
values by means of the 1991 dataset and then calculating the amplitudes
for the two 1991 subsets and the other datasets.  The frequency
analysis already gave some hint about the amplitude variability of some terms,
since for example  the 1.145~\cds peak was not clearly detected  in all datasets. 

This is confirmed by
the large amplitude variability, which can be discerned from  Tab.~\ref{lsq38}. 
Considering the 1991 subsets we can see differences of up to 2.3 mmag, but 
note the +3.7, +2.9 and --3.9 mmag differences between the amplitudes of 
the 1.627, 1.368 and 1.565~\cds terms in 1995 and 1994 datasets, which
look extremely large. The 
term ranking is also largely changing.  Note that the largest amplitude of the
1.565~\cds term was in 1994, whereas usually this term was below the first three in
1991 and 1995. The amplitude is also greatly changing.  The sum of the 
squared amplitudes is higher in the 1995 data (311 mmag$^2$) and lower in the
1994 dataset (250 mmag$^2$), with a variation of 25\%. This
difference supports the intrinsic variability of the mode amplitudes.
Note for example that in the 1991 data there are three terms having a
large amplitude, in 1994 data there are   four and in 1995 data  two
of them clearly have  the largest amplitude, considerably greater than the others.
Hence, the amplitude changes look more conspicuous in HD 224638 than in HD 224945. 

In the case of HD~224638 we can merge the 1991, 1994 and 1995 datasets
knowing that all the datasets show the same frequency content. Indeed, the
frequency analysis of the whole set yields the same frequencies listed in
Tab.~\ref{lsq38}.  The detection of the three terms
at 1.565, 1.627 and 1.697~\cds  supports the identification as three
independent modes proposed in  Sect.~5.1.
Unfortunately, we cannot give refined values for the frequencies,
owing to the cy$^{-1}$ aliases.
However, we performed some tests assuming a constant value
for the frequency  and then calculated the amplitudes and phases of each term
for each observing season. Besides the verification of the amplitude
variability, we found  that the same frequency displays similar phase values from one
season to the next. This fact supports an intrinsic variability  of the
mode amplitudes, rather than a beating phenomenon between two very close terms.

\section{Physical properties}
The properties of $\gamma$ Dor stars have been reviewed by Zerbi (2000);
driving mechanisms have been proposed by Guzik et al. (2000), Wu (2002) and
L\"offler (2002), but the origin of the pulsation in $\gamma$ Dor stars still remains
an open problem.

HD~224945 and HD~224638 have very similar physical properties
(Zerbi 2000):
$M_V$=2.98, $L/L_{\sun}$=5.5, $T_{\rm eff}$=7200~K, $R/R_{\sun}$=1.51 and
$M/M_{\sun}$=1.52 for HD~224638, $M_V$=3.07, $L/L_{\sun}$=5.1, $T_{\rm eff}$=7250~K, 
$R/R_{\sun}$=1.43 and $M/M_{\sun}$=1.51 for HD~224945. Also the difference in metallicity is
not significant, [Me/H]=--0.15 vs. --0.30, respectively. The two stars occupy very
similar positions also in the colour--magnitude diagram, being located on the ZAMS,
in the middle of the domain of $\gamma$ Dor stars and on the low temperature edge of
the $\delta$ Sct region (Handler 1999).
Therefore, two very similar
stars display  different pulsational modes, since HD~224945 is characterized
by four frequencies spread over a large interval, while HD~224638 displays a much 
more closely spaced set (it is similar to HR 2740; see Poretti et al. 1997). 
Moreover, the residual
power spectrum of HD~224638 has a very low noise level above 4.0~\cds, 
while that of HD~224945 has a higher level, suggesting some 
signal contribution.

The large spread in the frequency values observed for HD~224945 is another
piece of evidence that the cause of light variability in these stars is pulsation and
not stellar activity. Differential rotation is not able to match a
spread in frequencies as large as 0.7~\cd (considering the three terms always
observed, i.e., not considering the even more distant frequency at 1.16 \cd).

We note that multiperiodicity is not observed in all
$\gamma$ Dor stars.  Some display only a single photometric
period (i.e., a single low--order mode). 
A relevant example is HD~207223$\equiv$HR~8330 
(Aerts \& Kaye 2001), which is also monoperiodic
from a spectroscopic point of view (i.e. it does not show any high--order modes).
Another example is HD~164615 (Zerbi et al. 1997), one of the first known
$\gamma$ Dor stars.  It is probably a monoperiodic variable which shows amplitude
modulation.

Finally, we note that HD~224945 exhibits the shortest known period (i.e. 
0.3330~d) of the $\gamma$ Dor stars. Guzik et al. (2000) considered 0.48~d as
the lower limit (probably the old value of the same term, i.e. 2.00~\cd,
Paper~I). 
Therefore, the period value cannot be used to separate $p$-- and
$g$--modes, as $\delta$ Sct stars can display such `long--period' $p$--modes.
A careful evaluation of the physical parameters is necessary. Indeed, 
in HD~224638 and HD~224945 the fundamental radial mode is shorter than 0.07~d,
confirming that $p$-- and $g$--modes are well separated in $\gamma$ Dor stars.

\section{Conclusions}
The solution of the light curves of HD~224638 and HD~224945 has been deduced only on
the basis of a multisite campaign, since the true peaks could not be recognized if a
$\pm1~$\cds effect is present in the spectral window. 
We did our best to identify the correct excited
modes, but the more important characteristic of these two stars is the
strong multiperiodicity itself, independent of the exact frequency values.
In our opinion,  HD~224638 and HD~224945 
provide two of the best examples of features typical of  multiperiodicity
among $\gamma$ Dor variables:
different sets of frequency content, amplitude variations, disappearing terms, close
doublets of frequencies.  Also, after detecting five or six terms,
the rms scatter is larger than the observational error. Therefore, 
residual signal is hidden in the noise. 
Such extreme multiperiodicity can be clearly ascribed to
pulsation, as stellar activity is not able to generate it.

The differences in the frequency and amplitude
ranges of HD~224638 and HD~224945 (clearly visible in Fig.~\ref{lc})
remind us of the 
unpredictable frequency content of $\delta$ Sct stars,
where the selection mechanism among all the possible modes seems to be different
from one star to the next (Poretti 2000). 
The amplitude variability of the excited modes is another point of similarity
between $\gamma$ Dor and $\delta$ Sct stars. Other than multiperiodicity,
HD~224945 and HD~224638 provide the best
examples of amplitude variability among $\gamma$ Dor stars, following the 
three campaigns
carried out on these stars. This observational evidence suggests
strategies for  asteroseismological space missions: a long observing run (or 
two separate runs) may be very helpful in detecting more terms 
(excited at different levels
at different times) or in studying damping effects.

\begin{acknowledgements}
ER and SM acknowledge the partial support by 
the Junta de Andalucia and by the Direccion General de
Investigacion (DGI) under project AYA2000-1559. SM
also acknowledges the financial support by the Osservatorio
Astronomico di Brera and by the Agenzia Spaziale Italiana
(ASI Contract I/R/037/01). Thanks are due to an
anonymous referee and to M.~Breger for useful comments and
suggestions on the first version of the manuscript.
\end{acknowledgements}

\end{document}